\documentclass{article}

\usepackage{arxiv}

\usepackage[utf8]{inputenc} 
\usepackage[T1]{fontenc}    
\usepackage{hyperref}       
\usepackage{url}            
\usepackage{booktabs}       
\usepackage{amsfonts}       
\usepackage{nicefrac}       
\usepackage{microtype}      
\usepackage{lipsum}
\usepackage[utf8]{inputenc}
\usepackage{graphicx}

\title{Optionality and Convexity in ICT Networks}

\author{
  Edmond Shami\thanks{Edmond holds a Bsc in Telecommunications and Electronics Engineering from Jordan University of Science and Technology. He is also a Cisco CCNA and CCDA certified engineer and  has 3 years of experience in telecommunications design from Dar Al Handasah, where he is currently employed.} \\
  Telecom Design Engineer\\
  Dar Al-Handasah\\
  Amman, Jordan 11191 \\
  \texttt{edmond.shami@gmail.com} \\
   \
 \\
  \\
  \\
  \\
  \\
}

\begin{document}
\maketitle
\begin{abstract}
Networking companies, especially the ones with the biggest market shares, tend to offer “end-to-end” solutions for their customers, with large discounts, that it would sound irrational to decline such offers, helping contractors make larger profits, leaving clients more fragile to future uncertainties, and robbing the client from the leverage of optionality. This paper discusses the role of optionality in harnessing convex payoffs in uncertain domains, showing how ICT Networks' supply chain plays a big role in determining a better response to future developments, by harnessing the "multi-vendor" model, and ends with the story of how Wall Street made hundreds of millions using optionality in ICT Networks. 
\end{abstract}


\section{Introduction}
Optionality, is basically the ability to choose the best option out of multiple options. Which, in many cases, runs contrary to focusing your bets on fewer options by using forecasting and simulation models, to hedge your bets for "bigger payoffs" with a "minimum cost", forgetting the consequences of missing the actual best opportunity. In uncertain fields, such as ICT networks, where developments happen almost on a monthly basis, and demands for higher bandwidth and speeds increase nonlinearly, with standards in each area (wireless, active networks, structured cabling, storage, etc.) evolving very rapidly to accommodate technological developments to satisfy higher demand for bandwidth and speed, optionality plays a huge role, often overlooked by clients looking to reduce the initial cost of installation.

\section{Convex Payoffs through Optionality.}

In this section I will recall the role of optionality from [1].

In uncertain fields, convexity from optionality trumps knowledge (Figure 1).

\begin{figure}[htp]
    \centering
    \includegraphics[width=9cm]{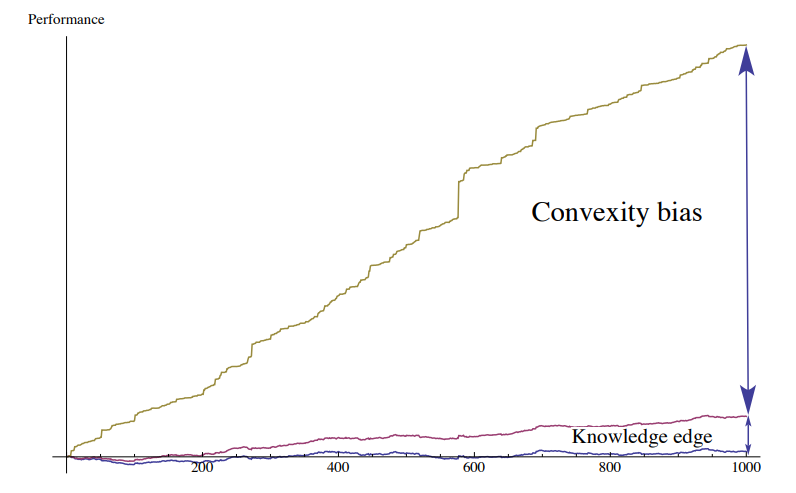}
    \caption{The Antifragility Edge. [1] }
    \label{fig: convexityknowledge}
\end{figure}

By distributing your bets (1/N), making N (N here represents the number of available options in the market that you can easily acquire/use) as big as possible, and reducing the costs per attempt, you can attain with minimal cost and risk, the best payoff (figure 2).

\begin{figure}[htp]
    \centering
    \includegraphics[width=9cm]{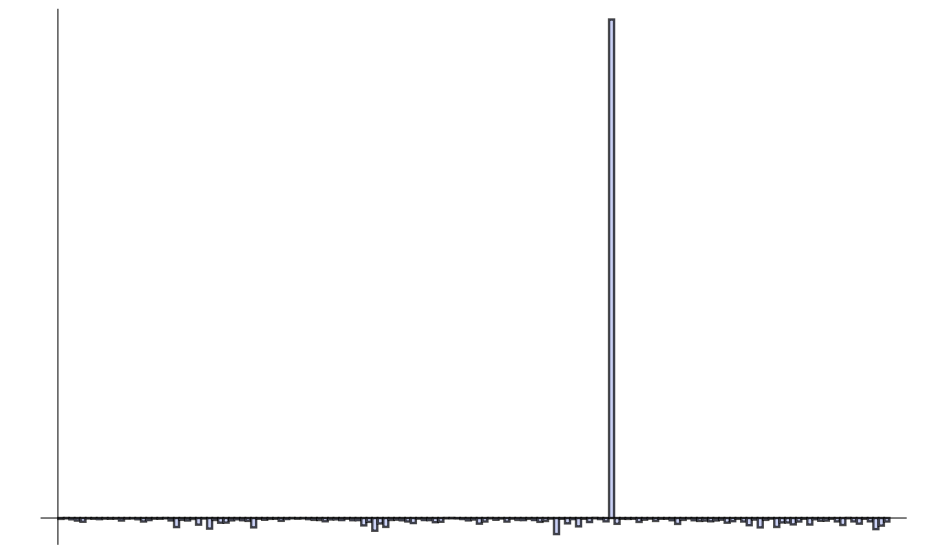}
    \caption{Small Probability, High Impact Payoffs: The horizontal line can be the payoff over time. [1]}
    \label{fig: payoff}
\end{figure}

And finally, as can be illustrated in figure 3, the higher the uncertainty, the higher the payoff. 

\begin{figure}[htp]
    \centering
    \includegraphics[width=15cm]{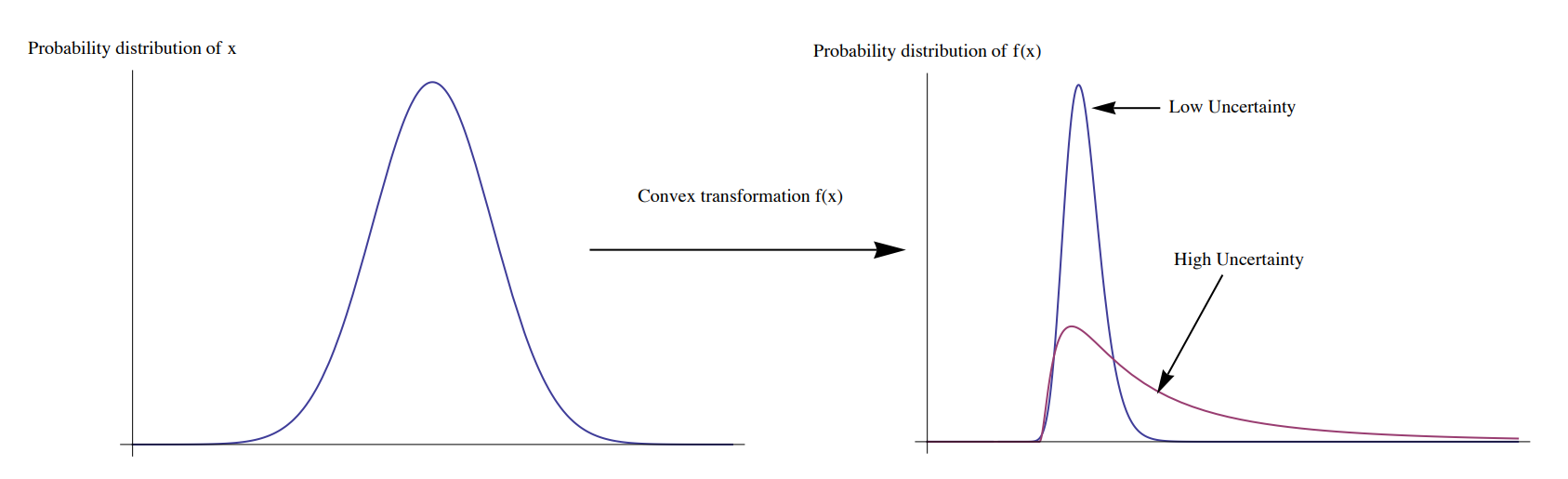}
    \caption{A convex transformation. Left is the symmetric distribution of outcomes for x. Right is asymmetric distribution of f(x). f(x) benefits from uncertainty: the more uncertainty, the higher the expected mean because negative events have no significant harm. [1]}
    \label{fig: convexity}
\end{figure}

\section{Optionality in Supply Chain Procurement for ICT Networks}

Now, how does optionality affect the supply chain procurement model you adopt for your network?

First, we apply a simple MILP optimization model, consisting of demand and supply parameters, where demand (D) is simply the needs of the network and the supply (S) is the capacity of the supply chain network to provide the best and lowest cost options, x represents the products you require and c the cost of each product, and z is the total cost.

Minimum of $z = \sum_{i} \sum_{j} c_{ij}  x_{ij} $

Where:

\begin{enumerate}
\item $\sum_{j} x_{ij} \leq S_{i}$ where                    $i \in S$ These are the supply constraints - the total number of shipped units from a supply node i to all demand nodes j must be less than (or equal to) the supply capacity of node i.
\item $\sum_{i} x_{ij} \geq D_{j}$ where                    $j \in D$ These are the demand constraints - the number of shipped units to a demand node j from all supply nodes i must be at least the demand at node j.
\item $x_{ij} \geq 0$ These are the non-negativity constraints.
\end{enumerate}

As can be inferred from this model simulation, increasing the constraints (reducing the number of suppliers) results in increasing the minimum optimized cost [2].

Second, item-supplier specific constraints are constraints that apply to one of the products at one of the suppliers (vendors). For instance, when procuring a specific product for the network, each supplier (vendor) may have capacity restrictions for a given product. In the formulation, these constraints are easily implemented as a separate matrix.  Not to forget that service attributes vary for individual items for each vendor [3].

Let’s take this point a bit further. Products usually fall under one of the following sections in terms of probability of disruption and the consequences [4]:

\begin{figure}[htp]
    \centering
    \includegraphics[width=9cm]{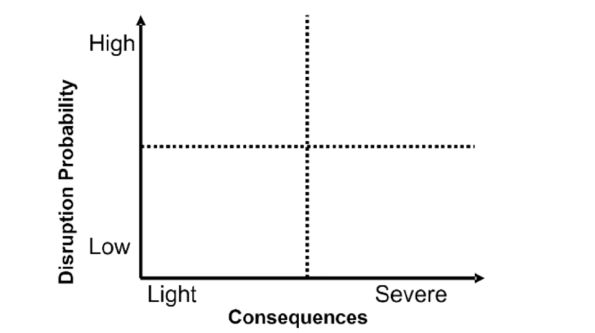}
    \caption{Disruption Probability vs Consequences}
    \label{fig: convexity}
\end{figure}
 
Assuming the probability of failure for a critical device is low, and network disruptions obey a power law: $f(x) =  x ^-k$, the product will fall under the following segment (figure 5).

\begin{figure}[htp]
    \centering
    \includegraphics[width=9cm]{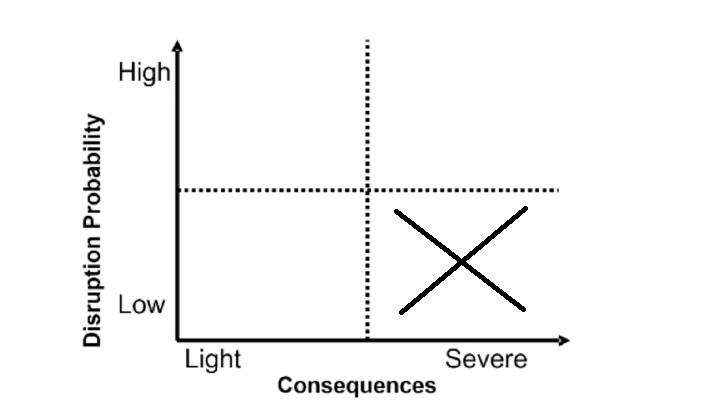}
    \caption{Low Disruption Probability and High Consequences}
    \label{fig: convexity}
\end{figure}

The procurement strategy in this case should have the following characteristics [3]:

\begin{enumerate}
\item Standardization 
\item Reduced transportation costs 
\item Global and active sourcing 
\end{enumerate}

And the above takes into consideration the client’s network only. But what about the disruptions that will affect a single supplier? These can include natural disasters, government and politics (trade barriers, political instability, trade embargo), and macroeconomic disruptions that causes economic contractions. Hence, if you rely on a single vendor, the disruption will put you at a huge risk as well.  

\section{A Case from Wall Street.}

Can optionality in networks design, help you make/lose millions?

Yes, and this can be illustrated by the rise of High Frequency Trading in the early 2000s. Where Wall Street companies started competing for milliseconds in network speed, replacing their network devices with faster devices, irrelevant of which company manufactured it. For if you were a few millisecond faster than your competitor, you can beat him to millions of dollars in profits on a single deal [5]. This is probably at the far end of the spectrum of uncertainty in the applications of ICT networks, but it can illustrate how implementing a multi-vendor network can have a better payoff than a single-vendor network.

\section{Conclusions and Recommendations}

From what has been argued above, you can notice that relying on a single vendor puts your network in a fragile position in terms of costs and future risks to the network. Hence, it is recommended to have your network compatible with as many vendors as possible, and to achieve that, you should:

\begin{enumerate}
\item Rely on international and industry standards and protocols (IEEE, IETF, etc.) for your network operations.
\item Implement redundancy in network components with 2 (at least) vendors. Such as having one spine switch using a specific vendor and the other spine using a different vendor.  
\item Segment the network to as many segments as possible. Where the need arises for a proprietary technology, it is only confined to that segment of the network. 
\end{enumerate}

ICT networks face a lot of uncertainty, and because our knowledge is always limited,  making early predictions of network performance and needs is not possible. That's why, using optionality is a key aspect of decision making in the early design stages of ICT networks.

\bibliographystyle{unsrt}  


\end{document}